\documentclass[aps,prl,twocolumn,groupedaddress]{revtex4}
\usepackage[pdftex]{graphicx}
\usepackage{latexsym}
\begin{document}

\title{ Influence of Nanoparticle Additives on the Fragility of \\ Polymer
  Glass Formation and the Buchenau Relation }

\author{\vspace{-2mm}Francis W. Starr}
\affiliation{Department of Physics, Wesleyan University,
  Middletown, CT 06459, USA} 
\author{\vspace{-4mm}Jack F. Douglas}
\affiliation{Polymers Division, National Institute of Standards and
  Technology, Gaithersburg, Maryland 20899, USA}

\date{submitted June 5, 2009}
 
\begin{abstract}
  
  We investigate the impact of the addition of nanoparticles (NP) on the
  fragility of a model glass-forming polymer melt by molecular dynamics
  simulations.  We find significant changes in fragility for
  nanoparticle volume fractions $\phi$ exceeding $\approx 5$~\%, where
  fragility changes correlate with the inverse variance of the magnitude
  of the Debye-Waller factor $\langle u^2\rangle$, a measure of local
  ``stiffness'' fluctuations. We also confirm the validity of the
  Buchenau relationship between $\langle u^2\rangle$ and the structural
  relaxation time $\tau$ for all $\phi$ and polymer-NP interaction
  types.
\end{abstract}

\maketitle

The addition of a small concentration of nanoparticles (NP) to
glass-forming polymer materials can lead to large property changes 
that are difficult to comprehend by
extension of the effects of macroscopic filler additives and the
corresponding theory of composite materials~\cite{am05,wypych}. For
well-dispersed NP, these changes are often rationalized by the large
surface-to-volume ratio of these particles, which results in a greater
interfacial interaction with the surrounding polymer matrix. However,
this is not a unique mechanism for all property changes, and other
mechanisms, such as chain bridging~\cite{schweizer,gersappe,kds07} and
NP self-assembly into extended structures~\cite{akcora09}, are under
current discussion.
 
Of the many affected properties, changes in the glass transition
temperature $T_g$ have been particularly emphasized, since changes in
$T_g$ are correlated with changes in diverse transport phenomena.
Specifically, both experimental and theoretical studies have indicated a
tendency for highly attractive or repulsive (non-attractive) polymer-NP
interactions to increase or decrease $T_g$, respectively.  This
phenomenon has been rationalized in terms of the influence of the NP
boundary interactions on the dynamics of polymers within an interfacial
layer near the NP surfaces~\cite{bansal05,fd01,ssg,tnp00}. In
particular, the interfacial polymer layer around the NP shows a slowing
down (increased $T_g$) or acceleration of dynamics (decreased $T_g$)
when the polymer-NP interactions are attractive or repulsive,
respectively. These observations led to the suggestion that the
interparticle distance (related to the particle concentration for
uniformly distributed NP) plays a role analogous to film thickness in
thin polymer films~\cite{ssg,bansal05}.

Changes of $T_g$, while informative, provide only a limited
understanding of how NP affect the properties of glass-forming polymer
melts. It is also natural to expect that the temperature dependence of
the dynamic properties approaching $T_g$, referred to as the
``fragility'' of glass formation~\cite{angell95,dfd}, will be altered.
Fragility changes have been argued for
on theoretical grounds~\cite{dfd}, based on the finding that any
factor that influences the molecular packing in the glass state ($T <
T_g$) should also alter the fragility of glass formation.  The NP we
study should be particularly effective at modifying molecular packing,
since their size is roughly commensurate with the heterogeneity scale of
fluids near their $T_g$ (i.e., 2~nm to 3~nm)~\cite{dyn-het}.

The present work addresses how polymer-NP interactions and NP
concentration affect the fragility of glass formation, and how fragility
changes relate to variations in the high frequency molecular dynamics,
as measured by the Debye-Waller factor $\langle u^2\rangle$.  We find
that the addition of nanoparticles can increase or decrease fragility,
depending on the polymer-NP interaction, and that this effect becomes
more pronounced with increasing $\phi$. We relate these changes to
fluctuations of $\langle u^2\rangle$, which can be interpreted as a
change in the fluctuations of the local molecular
stiffness~\cite{zaccai}. We also find that the Buchenau
relation~\cite{buchenau} holds with remarkable generality for all $\phi$
and interaction types, confirming the relationship between high
frequency relaxation and structural relaxation under rather general
circumstances.

Our findings are based on equilibrium molecular dynamics simulations of
a nanoparticle surrounded by a dense polymer melt, as well as
simulations of a pure melt for comparison purposes. We utilize periodic
boundary conditions so that our results correspond to a uniform
dispersion of NP.  The polymers are modeled by a well-studied
bead-spring model~\cite{fene2}.  All monomer pairs interact via a
Lennard Jones (LJ) potential $V_{LJ}$, and bonded monomers along a chain
are connected via a FENE anharmonic spring potential.  The NP consists
of 356 Lennard-Jones particles bonded to form an icosahedral NP; the
facet size of the NP roughly equals the equilibrium end-to-end distance
for a chain of 20 monomers.  Details of the simulation protocol and our
model potentials can be found in ref.~\cite{ssg}; further studies of the
clustering and mechanical properties related to this model are presented
in~\cite{sdg03,rds08}.

We simulate systems with 100, 200, or 400 chains of $M=20$ monomers each
(for totals of $N=2000$, 4000, and 8000 monomers) to address the effect
of varying the NP volume fraction.  Under constant pressure conditions,
the addition of nanoparticles can give rise to a change in the overall
melt density.  A slight change in density can cause a significant but
trivial change in the dynamic properties relative to the pure melt.  In
order to probe only changes caused by the interactions between the NP
and the polymer melt, we have matched the density of monomers far from
the NP with that of the pure polymer melt.

\begin{figure}[t]
\begin{center}
\includegraphics[clip,width=3.3in]{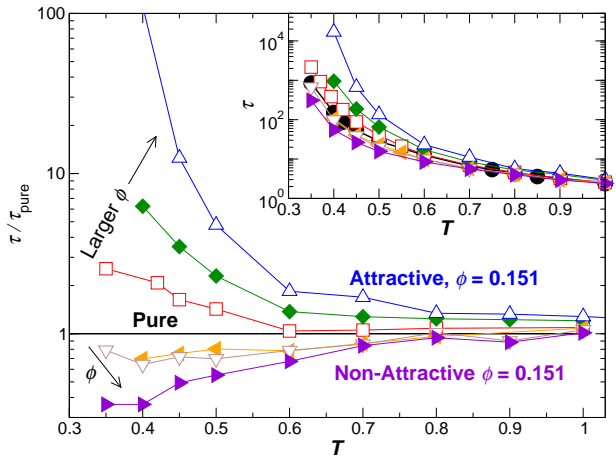}
\end{center}
\caption{Relaxation time $\tau$ as a function of $T$ for each $\phi$
  normalized by $\tau_{\rm pure}$ for the pure melt. The inset shows the
  raw data for $\tau$.  Attractive interactions lead to increases in
  $\tau$, while non-attractive interactions decrease $\tau$.  In both
  cases, the effect is more pronounced with increasing concentration.
  The concentrations are $\phi = 0.0426$ (red $\bigcirc$ and orange
  $\bigtriangleup$), 0.0817 (green $\Box$ and brown $\triangleleft$) and
  0.151 (blue $\Diamond$ and violet $\bigtriangledown$).  The pure melt
  is indicated in black.  The fits of the VFT form to
  the data deviate by at most 0.5~\%.\vspace{-5mm}}
\label{fig:tau}
\end{figure}

To quantify changes in $T_g$ and fragility, we evaluate effect of $\phi$
and the polymer-NP interactions on $\tau$, measured from the decay of
the coherent intermediate scattering function (see supplementary
information for a technical description).  The effects of these
interactions on $\tau$ for some $\phi$ were already presented in
ref.~\cite{ssg}; here we provide additional simulation data and expand
the analysis considered before to include the effect of the NP on
fragility.  As expected, Fig.~\ref{fig:tau} shows that attractive
polymer-NP interactions slow the relaxation ($\tau$ becomes larger),
while non-attractive polymer-NP interactions give rise to an increased
rate of relaxation ($\tau$ becomes smaller).  The effect of $\phi$ is
more clearly seen by rescaling $\tau$ by $\tau_{\rm pure}$ of the pure
melt, which shows that $\tau$ can be altered by a factor of more than an
order of magnitude on cooling.  The effect of the NP is more pronounced
at low $T$, since at high $T$ the polymer-NP interaction potential
strength is weak in comparison with the system kinetic energy.

\begin{figure}[t]
\begin{center}
\includegraphics[clip,width=3.3in]{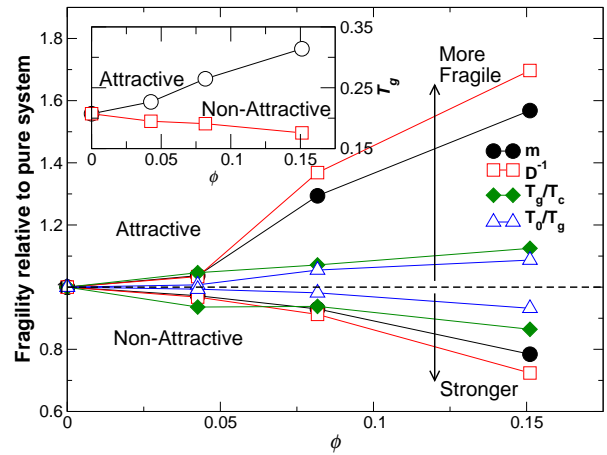}
\end{center}
\caption{Fragility dependence on $\phi$ relative to the pure
  melt.  We consider four different measures of fragility, which are
  discussed in the text.  All measures show the same qualitative trend:
  namely, the system with attractive polymer-NP interactions becomes
  more fragile, while the system with non-attractive interactions
  becomes less fragile (stronger).  The inset shows the changes in $T_g$
  for the different $\phi$ and interactions.\vspace{-5mm}}
\label{fig:fragility}
\end{figure}

We next examine how these changes in $\tau$ affect $T_g$ and fragility.
The inset of Fig.~\ref{fig:fragility} confirms that $T_g$ increases when
there is attraction, and decreases with non-attractive interactions. To
estimate $T_g$, we fit the data using the Vogel-Fulcher-Tammann (VFT)
expression~\cite{angell95}
\begin{equation}
\tau = \tau_0 e^{D/(T/T_0-1)}.
\label{eq:vft}
\end{equation}
$T_0$ is an extrapolated divergence temperature of $\tau$, while $D$
provides a measure of the fragility.  
We use the VFT fit to estimate $T_g$ based on the condition that
$\tau(T_g) = 100$~s (the canonical definition of the laboratory glass
transition~\cite{angell95}), assuming the one time unit in standard LJ
reduced units corresponds to 1~ps (reduced units are defined the in the
supplementary information).

Since $T_g$ changes have been the focus of previous work, we emphasize
how polymer-NP interactions and $\phi$ affect fragility.  Since there is
no single agreed upon measure of fragility, we consider several
different measures. First, as indicated above, the parameter $D$ from a
VFT fit to $\tau$ is widely utilized; specifically, a larger value of
$D$ indicates a stronger (less fragile) glass-forming fluid so that
$1/D$ increases with increasing fragility.  Another common estimate of
the fragility is defined by the temperature ratios $T_0/T_g$ or
$T_g/T_c$. 
We estimate $T_c$ using the power-law form $\tau \sim
(T/T_c-1)^{-\gamma}$ in an appropriate temperature range (see
supplementary information for fitting details). Since strong systems
should have relatively weak temperature dependence of $\tau$ approaching
$T_g$, larger values of $T_0/T_g$ or $T_g/T_c$ correspond to more
fragile systems.  Finally, we use the VFT fit to estimate the fragility
from the $T$ dependence of $\tau$ near $T_g$ using the most commonly
advocated fragility definition~\cite{sokolov},
\begin{equation}
m=d(\ln\tau)/d(T_g/T)|_{T_g}.
\end{equation}
For strong systems, the rate of change of $\tau$ with respect to $T$ is
smaller than that of fragile systems; hence $m$ is larger for more
fragile glass-forming fluids.  
We must rely on an extrapolation of the VFT fit to determine $m$, so
that caution should be exercised interpreting the precise values of our
$m$ estimates.

We summarize the results for the various fragility metrics in
Fig.~\ref{fig:fragility}, where we find that attractive polymer-NP
interactions lead to more fragile glass formation as a function of
$\phi$; conversely, non-attractive polymer-NP interactions lead to
stronger glass formation.  These changes are non-trivial, since $T_g$ is
independent or anticorrelated with fragility in some
systems~\cite{mckenna}.  Our findings for the changes of fragility are
consistent with several experimental studies.  Bansal et
al.~\cite{bansal05} found that dispersions of NP having repulsive
interactions caused $T_g$ to decrease, accompanied by an appreciable
broadening of the glass transition region.  Though they did not
interpret their results in terms of fragility, the increased breadth is
indicative of increased strength (decreased fragility), as we observe.
For the case of fullerenes dispersed in polystyrene, Sanz et
al.~\cite{sanz} reported behavior expected for attractive polymer-NP
interactions, namely an increase in $T_g$, accompanied by an increased
fragility. Other studies have indicated no detectable or only a small
change in the fragility with NP additive for small
$\phi$~\cite{green}. The apparent absence of fragility changes at small
$\phi$ is also consistent with our results, which likewise show nearly
undetectable changes in this concentration range.

\begin{figure}[t]
\begin{center}
\includegraphics[clip,width=3.3in]{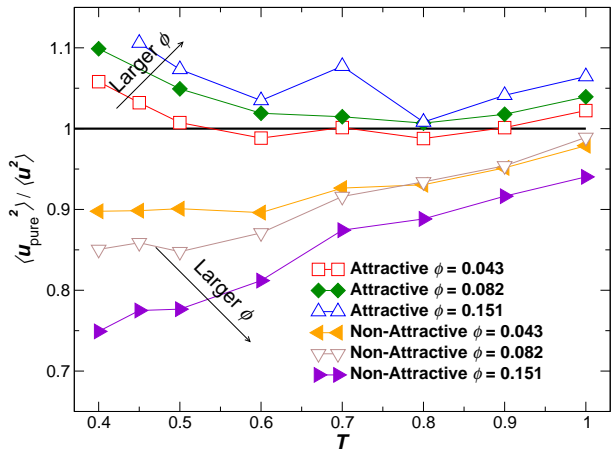}
\end{center}
\caption{$T$ dependence of $1/\langle u^2 \rangle$ for different $\phi$
  normalized by $1/\langle u^2 \rangle$ for the pure melt. Note the
  parallelism to fig.~\ref{fig:tau}.\vspace{-5mm}}
\label{fig:dw-norm}
\end{figure}

Having established the variation in $T_g$ and fragility for the various
systems, we now examine how these changes can be understood from the
high frequency melt dynamics.  Both experiments and simulations have
shown that the Debye-Waller factor $\langle u^2\rangle$ can be related
to the low frequency relaxation associated with large scale structural
relaxation~\cite{ssdg02,larini}.  $\langle u^2\rangle$ measures monomer
displacement on a time scale over which the particles are caged by their
neighbors, and is accessible from both x-ray and neutron scattering
measurements~\cite{dw-exp}.  Since $\langle u^2\rangle$ is usually
determined experimentally at a fixed instrumental time corresponding to
the time scale on the order of vibrational motion of the molecules, we
determine the mean-squared chain segment displacement (MSD) at a
vibrational time scale, specifically at a reduced time equal to 1.53,
$\approx$ the mean collision time.

Fig.~\ref{fig:dw-norm} shows that, when normalized by the behavior of
the pure melt, the changes in $\langle u^2\rangle_{\rm pure}/ \langle
u^2\rangle$ are very similar to those observed for $\log (\tau/\tau_{\rm
  pure})$.  Evidently, the $\langle u^2\rangle$ is sensitive not only to
changes in $\phi$, but also to the polymer-NP interactions.  This
affirms that $\langle u^2\rangle$ is a potentially useful indicator of
changes in low frequency relaxation.

\begin{figure}[t]
\begin{center}
\includegraphics[clip,width=3.3in]{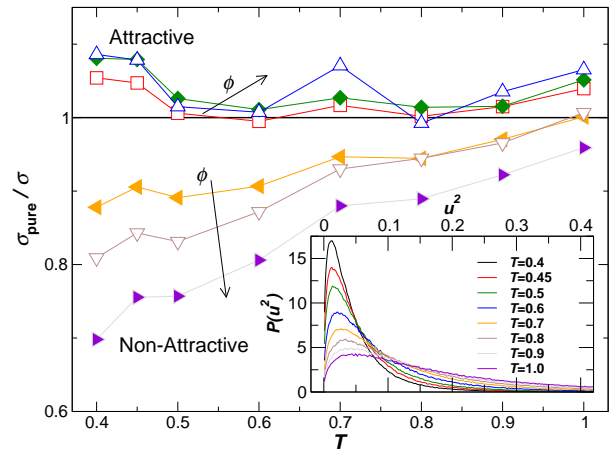}
\end{center}
\caption{$T$ dependence of the inverse variance $1/\sigma$ associated
  with $\langle u^2 \rangle$ for different concentrations
  normalized by $\sigma$ for the pure melt. Symbols are the same as
  figs.~\ref{fig:tau} and \ref{fig:dw-norm}. The changes in $\sigma$
  mirror the changes of fragility, as argued in the text.
  The inset of the figure provides a reference for the $T$ dependence of
  the distribution $P(u^2)$ for the system with attractive polymer-NP
  interactions.\vspace{-5mm}
}
\label{fig:dw-variance-norm}
\end{figure}

To understand how the changes in fragility relate to the high frequency
dynamics, we must look beyond changes in $\langle u^2\rangle$. It has
been argued~\cite{dfd,rydp} that the changes in fragility can be
connected with the efficiency of local packing,
which is accompanied by changes in the fluctuations
in the local moduli.  Since the {\it inverse} of $\langle u^2 \rangle$ is a
measure of the local liquid rigidity at high frequency~\cite{zaccai}, 
we expect that fragility changes should be reflected in the variance
$\sigma^2 = \langle u^4 \rangle - \langle u^2 \rangle^2$ of the
Debye-Waller factor.

Figure~\ref{fig:dw-variance-norm} shows that, when normalized by the
value for the pure melt, $\sigma$ is larger for stronger glass formation
(the non-attractive interactions), and smaller in the more fragile case
(attractive interactions) in the $T$ range considered.  If we consider
the normalized inverse of this quantity $\sigma_{\rm pure}/\sigma$, we
can examine to the relative variance in the local stiffness.  These
results show that the stronger systems have relatively smaller
fluctuations in the local stiffness, while the more fragile systems have
relatively larger fluctuations.  This is consistent with the idea that
the the stronger systems should be better packed, and vice-versa for the
more fragile systems.  Hence, $1/\sigma$ appears to be reflective of
a change in fragility.

\begin{figure}[t]
\begin{center}
\includegraphics[clip,width=3.3in]{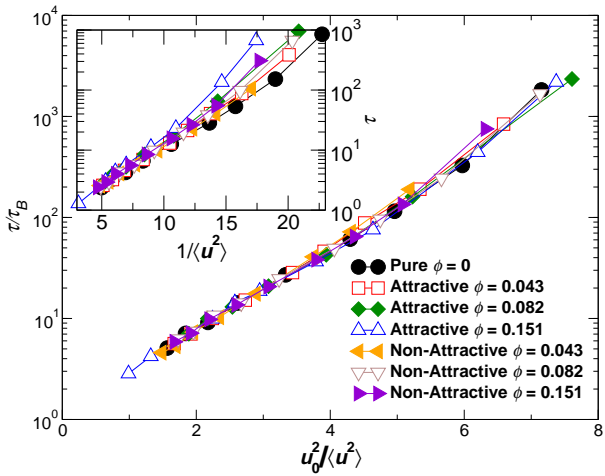}
\end{center}
\caption{Parametric relation between $\tau$ and $\langle u^2\rangle$.
  The inset shows the raw data, and the main plot shows the data
  collapse when the axes are scaled using the constants of the Buchenau
  relation.\vspace{-5mm}}
\label{fig:buchenau}
\end{figure}

Finally, we explore the proposed quantitative relation between $\tau$ and
$\langle u^2\rangle$ provided by the ``Buchenau relation,''
\begin{equation}
\tau = \tau_B e^{u_0^2/\langle u^2 \rangle},
\label{eq:buchenau}
\end{equation}
where $\tau_B$ and $u_0^2$ are system dependent constants.
This relation has been verified for a number of systems,
including the pure polymer melt currently under
investigation~\cite{ssdg02}.  Larini et al.~\cite{larini} have shown
that a generalization of this relation seems to hold for diverse
collection of simulated and real fluids over a wide range of
temperatures, suggesting an amazing generality of the Buchenau
expression, supporting the proposal that $\tau$ is a universal function
of $\langle u^2 \rangle$.  The inset in Fig.~\ref{fig:buchenau} shows a
striking agreement of our relaxation data with the Buchenau relation,
the correlation becoming better at lower $T$. Deviations at high $T$ can be
expected since $\langle u^2\rangle$ becomes progressively ill-defined as
$T$ increases.  More significantly, if we reduce $\tau$ and $\langle u^2
\rangle$ by the corresponding fit parameters $\tau_B$ and $u_0^2$, we
find that the data for all $\phi$ and $T$ can all be collapsed to a
single master curve (main part of figure).  This reinforces the somewhat
surprising fact that the slow dynamics of the system are intimately
connected with the high frequency vibrational properties of the system,
as epitomized by $\langle u^2\rangle$.

Similar to our findings, Riggleman et al.~\cite{rydp} have shown that
changes in fragility associated with the addition of molecular additives
are linked to changes in the shear modulus of the material, another high
frequency dynamical property. Additionally, Papakonstantopoulos et
al.~\cite{papa05} have suggested a link between fragility variations and
elastic constant (shear modulus) fluctuations in polymer nanocomposites
in the glass state, but this was not verified through a direct
determination of fragility.

Care must be taken in comparing the glass below $T_g$ with a liquid
approaching $T_g$ from above, as recent results suggest opposite trends
may result.  For example, Riggleman et al.~\cite{rydp} show that a
decrease of the fragility actually results in a stiffening below $T_g$,
but a softening above $T_g$, consistent with our results above $T_g$.
Such an inversion between relative material stiffness and relaxation
times near $T_g$ has been suggested to be a general
phenomenon~\cite{tatiana08}, having potential applications in relation
to scratch resistance of films, stabilizing nano-fabricated structures
formed by lithography and nano-imprint technology, and the preservation
of drugs in glass-forming preservative formulations~\cite{tatiana08}.

\end{document}